\documentclass[letterpaper]{ptephy_v1}

\preprintnumber{xxxx-xxxx} 

\begin{document}

\title{Search for the deeply bound $K^-pp$ state from the semi-inclusive forward-neutron spectrum in the in-flight $K^-$ reaction on helium-3}

\newcommand{\utokyo}{1}
\newcommand{\rcnp}{2} \newcommand{\victoria}{3} \newcommand{\seoul}{4} \newcommand{\infnhh}{5} \newcommand{\smi}{6} \newcommand{\torino}{7}
\newcommand{\torinoU}{8} \newcommand{\frascati}{9} \newcommand{\kyoto}{10} \newcommand{\osakaE}{11} \newcommand{\kek}{12} \newcommand{\osakaU}{13}
\newcommand{\riken}{14}
\newcommand{\titec}{15} \newcommand{\tum}{16} \newcommand{\komaba}{17} \newcommand{\tohoku}{18} \newcommand{\ecutum}{19} \newcommand{\kirams}{20} 
\author[\utokyo,*]{\collaborator{J-PARC E15 Collaboration}T. Hashimoto \thanks{Present address: RIKEN Nishina Center, RIKEN, Wako, 351-0198, Japan}}

\affil[\utokyo]{Department of Physics, The University of Tokyo, Tokyo, 113-0033, Japan }
\affil[\rcnp]{Research Center for Nuclear Physics (RCNP), Osaka University, Osaka, 567-0047, Japan }
\affil[\victoria]{Department of Physics and Astronomy, University of Victoria, Victoria BC V8W 3P6, Canada }
\affil[\seoul]{Department of Physics, Seoul National University, Seoul, 151-742, South Korea }
\affil[\infnhh]{National Institute of Physics and Nuclear Engineering - IFIN HH, Romania }
\affil[\smi]{Stefan-Meyer-Institut f\"{u}r subatomare Physik, A-1090 Vienna, Austria }
\affil[\torino]{INFN Sezione di Torino, Torino, Italy }
\affil[\torinoU]{Dipartimento di Fisica Generale, Universita' di Torino, Torino, Italy }
\affil[\frascati]{Laboratori Nazionali di Frascati dell' INFN, I-00044 Frascati, Italy }
\affil[\kyoto]{Department of Physics, Kyoto University, Kyoto, 606-8502, Japan }
\affil[\osakaE]{Laboratory of Physics, Osaka Electro-Communication University, Osaka, 572-8530, Japan }
\affil[\kek]{High Energy Accelerator Research Organization (KEK), Tsukuba, 305-0801, Japan }
\affil[\osakaU]{Department of Physics, Osaka University, Osaka, 560-0043, Japan }
\affil[\riken]{RIKEN Nishina Center, RIKEN, Wako, 351-0198, Japan }
\affil[\titec]{Department of Physics, Tokyo Institute of Technology, Tokyo, 152-8551, Japan }
\affil[\tum]{Technische Universit\"{a}t M\"{u}nchen, D-85748, Garching, Germany }
\affil[\komaba]{Graduate School of Arts and Sciences, The University of Tokyo, Tokyo, 153-8902, Japan }
\affil[\tohoku]{Department of Physics, Tohoku University, Sendai, 980-8578, Japan }
\affil[\ecutum]{Excellence Cluster Universe, Technische Universit\"{a}t M\"{u}nchen, D-85748, Garching, Germany }
\affil[\kirams]{Korea Institute of Radiological and Medical Sciences (KIRAMS), Seoul, 139-706, South Korea \email{tadashi.hashimoto@riken.jp}}
\author[\rcnp]{S. Ajimura}
\author[\victoria]{G. Beer}
\author[\seoul]{H.~Bhang}
\author[\infnhh]{M.~Bragadireanu}
\author[\torino,\torinoU]{L.~Busso}
\author[\smi]{M.~Cargnelli}
\author[\seoul]{S.~Choi}
\author[\frascati]{C.~Curceanu}
\author[\rcnp]{S.~Enomoto}
\author[\torino,\torinoU]{D.~Faso}
\author[\kyoto]{H.~Fujioka}
\author[\utokyo]{Y.~Fujiwara}
\author[\osakaE]{T.~Fukuda}
\author[\frascati]{C.~Guaraldo}
\author[\utokyo]{R.~S.~Hayano}
\author[\rcnp]{T.~Hiraiwa}
\author[\kek]{M.~Iio}
\author[\frascati]{M.~Iliescu}
\author[\osakaU]{K.~Inoue}
\author[\kyoto]{Y.~Ishiguro}
\author[\utokyo]{T.~Ishikawa}
\author[\kek]{S.~Ishimoto}
\author[\riken]{K.~Itahashi}
\author[\kek]{M.~Iwai}
\author[\riken,\titec]{M.~Iwasaki}
\author[\riken]{Y.~Kato}
\author[\osakaU]{S.~Kawasaki}
\author[\tum]{P.~Kienle\thanks{deceased}}
\author[\titec]{H.~Kou}
\author[\riken]{Y.~Ma}
\author[\smi]{J.~Marton}
\author[\komaba]{Y.~Matsuda}
\author[\osakaE]{Y.~Mizoi}
\author[\torino]{O.~Morra}
\author[\kyoto]{T.~Nagae}
\author[\rcnp]{H.~Noumi}
\author[\riken,\rcnp]{H.~Ohnishi}
\author[\riken]{S.~Okada}
\author[\riken]{H.~Outa}
\author[\frascati]{K.~Piscicchia}
\author[\frascati]{M.~Poli~Lener}
\author[\frascati]{A.~Romero~Vidal}
\author[\kyoto]{Y.~Sada}
\author[\osakaU]{A.~Sakaguchi}
\author[\riken]{F.~Sakuma}
\author[\riken]{M.~Sato}
\author[\frascati]{A.~Scordo}
\author[\kek]{M.~Sekimoto}
\author[\frascati]{H.~Shi}
\author[\frascati,\infnhh]{D.~Sirghi}
\author[\frascati,\infnhh]{F.~Sirghi}
\author[\kek]{S.~Suzuki}
\author[\utokyo]{T.~Suzuki}
\author[\seoul]{K.~Tanida}
\author[\utokyo]{H.~Tatsuno}
\author[\titec]{M.~Tokuda}
\author[\kyoto]{D.~Tomono}
\author[\kek]{A.~Toyoda}
\author[\tohoku]{K.~Tsukada}
\author[\frascati,\ecutum]{O.~Vazquez~Doce}
\author[\smi]{E.~Widmann}
\author[\osakaU]{T.~Yamaga}
\author[\utokyo,\riken]{T.~Yamazaki}
\author[\kirams]{H.~Yim}
\author[\riken]{Q.~Zhang}
\author[\smi]{J.~Zmeskal}






\begin{abstract}%
An experiment to search for the $K^-pp$ bound state was performed via the in-flight $^3$He($K^-,n)$ reaction using 5.3 $\times$ $10^9$ kaons at 1 GeV/$c$ at the J-PARC hadron experimental facility. In the semi-inclusive neutron missing-mass spectrum at $\theta_{n}^{lab}=0^\circ$, no significant peak was observed in the region corresponding to $K^-pp$ binding energy larger than 80 MeV, where a bump structure has been reported in the $\Lambda p$ final state in different reactions. Assuming the state to be isotropically decaying into $\Lambda p$, mass-dependent upper limits on the production cross section were determined to be 30--180, 70--250, and 100--270 $\mu$b/sr, for the natural widths of 20, 60, and 100 MeV, respectively, at 95\% confidence level. 
\end{abstract}

\subjectindex{D01, D33}

\maketitle

\section{Introduction}
The existence of a strongly-attractive force between antikaons ($\bar K$) and nucleons in isospin 0 channels leads to the prediction of the formation of deeply bound kaonic-nuclei \cite{PhysRevC.65.044005,Yamazaki:2002wp}. The investigation of those exotic states will provide unique information to reveal the sub-threshold $\bar KN$ interaction, which cannot be directly probed either by x-ray measurements \cite{PhysRevLett.78.3067,PhysRevLett.94.212302,Bazzi:2011wva} or by the low-energy $\bar KN$ scattering experiments \cite{Martin:1980qe}. The properties of kaonic nuclei are also of great interest, since they might open a doorway to high-density nuclear matter \cite{PhysRevC.65.044005,Dote:2004cj,Dote:2004ht}. However, their existence has not been conclusively established to date. 

The simplest kaonic nucleus is theoretically considered to be the so-called $K^-pp$ state\cite{Yamazaki:2002wp}; more generally, it is expressed as [$\bar K \otimes\{NN\}_{I=1,S=0}]_{I=1/2}$ with $J^\pi$=$0^{-}$. 
Intensive theoretical works based on a few-body calculation have been performed for the $K^-pp$ system, all of which predicted the existence of bound states \cite{Yamazaki:2002wp,Shevchenko:2007ie,Wycech:2009tg,Dote:2009um,Ikeda:2007vh,Ikeda:2010bd,Barnea:2012gk,MAEDA:2013ey,Bayar:2013ft}. However, the predicted binding energies (B.E.) and widths ($\Gamma$) are widely spread from 9--95 MeV and 34--110 MeV, respectively, primarily depending on the $\bar KN$ interaction models. 

On the experimental side, the FINUDA collaboration at DA$\Phi$NE investigated the stopped $K^-$ reaction on $^6$Li, $^7$Li, and $^{12}$C, and observed a bump structure in the invariant-mass spectrum of back-to-back $\Lambda p$ pairs \cite{PhysRevLett.94.212303}. They determined the B.E. and $\Gamma$ to be $115^{+6}_{-5} ({\it stat.}) ^{+3}_{-4}({\it syst.})$ MeV and $67^{+14}_{-11} ({\it stat.}) ^{+2}_{-3} ({\it syst.})$ MeV, respectively, although an alternative explanation for the bump was given in Ref. \cite{Magas:2006kh} in terms of $K^-$ absorption by nucleon pairs followed by re-scattering. The DISTO collaboration at SATURNE analyzed their dataset of the exclusive $pp\to\Lambda p K^+$ channel and observed a bump structure in the $K^+$ missing-mass and the $p\Lambda$ invariant-mass spectra at $T_p$=2.85 GeV \cite{Yamazaki:2010vx}, with a large cross section comparable to $\Lambda(1405)$ production, as predicted \cite{Yamazaki:2007vr}. The B.E. and $\Gamma$ were determined to be 103 $\pm$ 3 ({\it stat.}) $\pm$ 5 ({\it syst.}) MeV and 118 $\pm$ 8 ({\it stat.}) $\pm$ 10 ({\it syst.}) MeV, respectively. 
They also reported the absence of the peak structure at $T_p$=2.5 GeV, where the formation of $\Lambda(1405)$ drops down \cite{Yamazaki:2007vr,Kienle:2012dm}.
The HADES collaboration at GSI used the same reaction as DISTO at $T_p$=3.5 GeV. Their partial wave analysis of $pK^+\Lambda$ events determined upper limits on the production cross section of the $K^-pp$ state, which are smaller than that of $\Lambda$(1405) \cite{Agakishiev:2014ut}. In addition to the above, a less significant signal was reported in the stopped-$\bar p$ reaction \cite{Bendiscioli:2007tn} and no peak was observed in the $\gamma$-induced reaction \cite{Tokiyasu:2013mwa}. A $d(\pi^+,K^+)$ reaction study was also performed recently (J-PARC E27). In their proton coincidence analysis, a broad bump structure corresponding to the $K^-pp$ binding energy of $95 \pm 18 (stat.) \pm 30 (syst.)$ MeV and the width of $162 \pm 87 (stat.) \pm 66 (syst.)$ MeV was reported \cite{Ichikawa:2015ec}.

Since the $K^-pp$ is a basic ingredient for kaonic nuclear bound states, it is important to investigate it using different methods. For further investigation, we chose the in-flight kaon-induced reaction on helium-3 to observe a neutron at $\theta_n^{lab}= 0^\circ$:
\begin{eqnarray*}
K^- + {\rm ^3He} \to K^-pp + n.
\end{eqnarray*} 
The kaon beam momentum was selected to be 1 GeV/$c$, based on the $K^-$ beam yield and the elementary $K^-N$ reaction rate \cite{Beringer:2012df}. In this reaction, the momentum transfer is relatively small (0.2--0.4 GeV/$c$), and the forward-going neutron has a momentum of 1.2--1.4 GeV$/c$, so relatively large sticking probability could be expected. The kinematics of the in-flight reaction at a forward angle discriminates against the severe physics background encountered in the stopped $K^-$ experiment at KEK \cite{Sato2008107,Yim:2010ed}, namely, 1) neutrons emitted from hyperon decay, $Y\to n\pi$, and 2) hyperon production via the non-mesonic two-nucleon absorption processes, $K^-NN\to YN$. In contrast to the stopped $K^-$ reaction, those processes do not overlap in the deeply-bound region (B.E.=100--200 MeV) in the missing-mass spectrum of the forward neutron for the incident kaon momentum of 1 GeV/$c$. Another advantage of an in-flight $K^-$ experiment is that a quantitative analysis of the production cross section is possible, by comparing with the yield of the quasi-elastic and charge-exchange reactions; $K^-N\to K^-N$ and $K^-N\to\bar K^0N$. The effectiveness of these advantages of the in-flight method has already been demonstrated by the previous experiment using rather heavy targets \cite{Kishimoto2005383,Kishimoto:2007kr}. The present measurement on a helium-3 target is the first attempt to apply the in-flight ($K^-,n$) method for the $K^-pp$ system. 

There are two calculations of the spectral function for the forward neutron in this reaction. Koike and Harada predict a large cross section for the $K^-pp$ bound state, as much as 2--3 mb/sr, at $\theta_n^{lab} = 0^\circ$ using several phenomenological $\bar KN$ potentials \cite{Koike:2009cx}. With a potential which reproduces the binding energy and width reported by the FINUDA and DISTO collaborations, a distinct peak structure appears in the $^3$He$(K^-,n)X$ missing mass spectrum. Another theoretical spectral function was calculated by Yamagata-Sekihara {\it et al.} based on a potential derived in the framework of a chiral-unitary model \cite{YamagataSekihara:2009bw}. They predict a broad, loosely-bound observable state with a few hundreds of $\mu$b/sr cross section. 

In this Letter, the results of the first measurement of the in-flight $^3$He($K^-,n)$ reaction are reported. The present results were obtained based on a dataset of the J-PARC E15 experiment performed in May, 2013 with 5.3 $\times$ $10^9$ incident kaons on a helium-3 target.

\section{Experiment and analysis}
\begin{figure}
\begin{center}
\includegraphics[width=0.8\columnwidth]{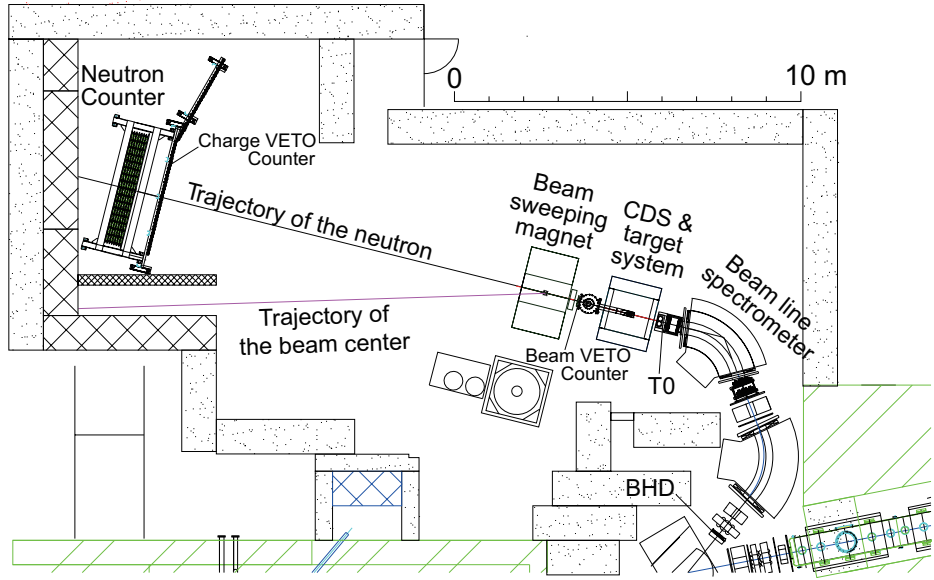}
\caption{\label{fig-k18br}Schematic view of the K1.8BR spectrometer\cite{Agari:2012vt}. The apparatus consists of a beam line spectrometer, a cylindrical detector system (CDS) that surrounds the liquid helium-3 target system to detect the decay particles from the target region, a beam-sweeping magnet, and a neutron time-of-flight counter located $\sim$ 15 m downstream from the target position.}
\end{center}
\end{figure}

The J-PARC E15 experiment was performed at the K1.8BR beam line of the J-PARC hadron experimental facility \cite{Agari:2012uu}. The experimental apparatus is schematically shown in Fig.~\ref{fig-k18br} and is briefly discussed below. More detailed description can be found elsewhere \cite{Agari:2012vt}.

A $K^-$ beam, purified with an electrostatic separator, was identified with an aerogel Cherenkov counter and was confirmed by an offline time-of-flight analysis between a beam-line hodoscope (BHD) and a time-zero counter (T0). The beam momentum was analyzed with a beam-line spectrometer consisting of two sets of drift chambers installed across a dipole magnet. The momentum resolution was (2.0 $\pm$ 0.5) $\times\ 10^{-3}$ with an absolute precision of 2 MeV/$c$ at 1 GeV/$c$. A typical $K^-$ yield was 1.5 $\times\ 10^5$ per spill\footnote{30 $\times\ 10^{12}$ primary protons at 30 GeV on a 50\%-loss Au production target. The spill length was $\sim2$ seconds with a 6-second repetition cycle.} with a $K^-/\pi^-$ ratio of 0.45. 

A cylindrical target cell, 137 mm long and 68 mm in diameter, was filled with liquid helium-3, and placed at the final focus point of the beam line. The density of the target was 0.081 g/cm$^3$ at a temperature of 1.4 K.  

The momentum of the forward neutron was measured by the time-of-flight method between T0 and a plastic scintillation counter array called neutron counter (NC), placed at a distance of $\sim$ 15 m from the target. The NC, segmented into 16-column (horizontal) $\times$ 7-layer (depth) units with a total volume of 3.2 m (horizontal) $\times$ 1.5 m (height) $\times$ 0.35 m (depth), had a coverage of 22 msr solid angle. Charged particles along the beam line were swept out from the NC acceptance by a beam sweeping magnet. Furthermore, two types of scintillation counter arrays were installed, downstream of the target system and upstream of the NC, to veto charged particles.

To determine the flight length of a forward-going particle, the reaction vertex was reconstructed by a cylindrical detector system (CDS) surrounding the target. Charged particles emitted in the reaction were tracked with a 15-layer cylindrical drift chamber (CDC) in a 0.7 T solenoidal field. To reconstruct the reaction vertex, a kaon track reconstructed by a drift chamber placed just in front of the target was utilized together with the CDC tracks. The vertex resolutions ($\sigma$) were $\sim$ 1 and $\sim$ 7 mm in the perpendicular and parallel directions to the beam, respectively. The helium-3 fiducial volume was defined as 60 mm in diameter and 100 mm in length to remove the events where beam kaons interact with the target cell. A cylindrical hodoscope (CDH) was used as a trigger counter with a polar-angle acceptance from 54 to 126 degrees, corresponding to a solid angle coverage of 59\% of 4$\pi$. The timing of the CDH provided particle identification together with the track momentum analyzed by the CDC. The typical CDS momentum acceptance of 80, 180, and 260 MeV/$c$ for pions, kaons, and protons, respectively, is limited by the materials between the target and the CDH. 

\begin{figure}
\begin{center}
\includegraphics[width=0.7\columnwidth]{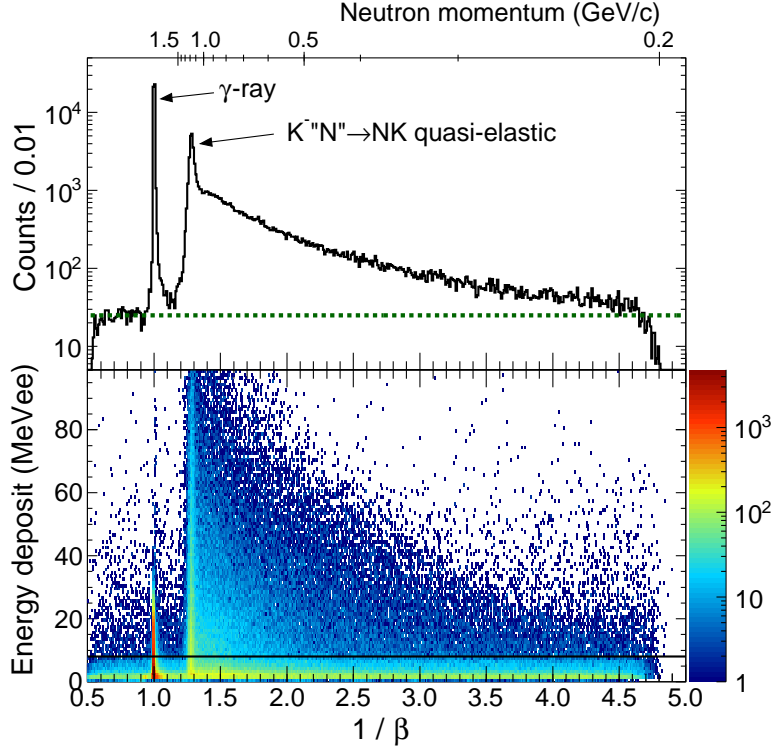}
\caption{(top) 1/$\beta$ distribution based on an off-line threshold. The dotted line indicates the accidental background level evaluated in the region from 0.6 to 0.9. (bottom) Distribution of the energy deposited on the NC versus 1/$\beta$. The horizontal solid line indicates the 8 MeV$ee$ offline threshold applied to reduce the accidental background.  \label{fig-beta}}
\end{center}
\end{figure}

Figure~\ref{fig-beta} shows a 1/$\beta$ spectrum of forward neutral particles detected by the NC, and the measured energy deposited versus 1/$\beta$. The spectrum was obtained based on a {\it semi}-inclusive condition by requiring at least one charged track in the CDS to reconstruct the reaction vertex. The $\gamma$-ray peak position provided a reference to define the time zero and to correct time-walk effects of each NC segment. The time-of-flight resolution for the forward neutral particles was obtained from the width of the $\gamma$-ray peak to be 150 ps ($\sigma$). Another peak at around $1/\beta=1.3$ comes from the quasi-elastic scattering ($K^-``n" \to K^-n$: double quotation marks indicate a quasi-free nucleon in $^3$He) and the charge-exchange reaction ($K^-``p" \to K^0_sn$). Here the neutron timing was determined by the time-wise first-hit segment with an energy deposit larger than a threshold determined offline. The threshold was optimized to be 8 MeV$ee$ in terms of a signal-to-background ratio at the quasi-elastic peak. The accidental background can be evaluated from the unphysical region, $1/\beta<1$ where causality does not hold, as the dotted line in Fig.~\ref{fig-beta}. The signal-to-background ratio at the quasi-elastic peak was $\sim$ 100. The detection efficiency for a neutron obtained was 0.23 $\pm$ 0.04 using $\sim$ 1.1 GeV/$c$ neutrons by an exclusive analysis of the $^3$He($K^-,nK^0_s)d$ reaction. We assume the neutron detection efficiency has no momentum dependence around 1 GeV/$c$ since $np$ and $n$C reaction cross sections are known to be flat in this region \cite{jendl}.

\section{Results and Discussion}
\subsection{Semi-inclusive neutron spectrum}
\begin{figure}
\begin{center}
\includegraphics[width=0.7\columnwidth]{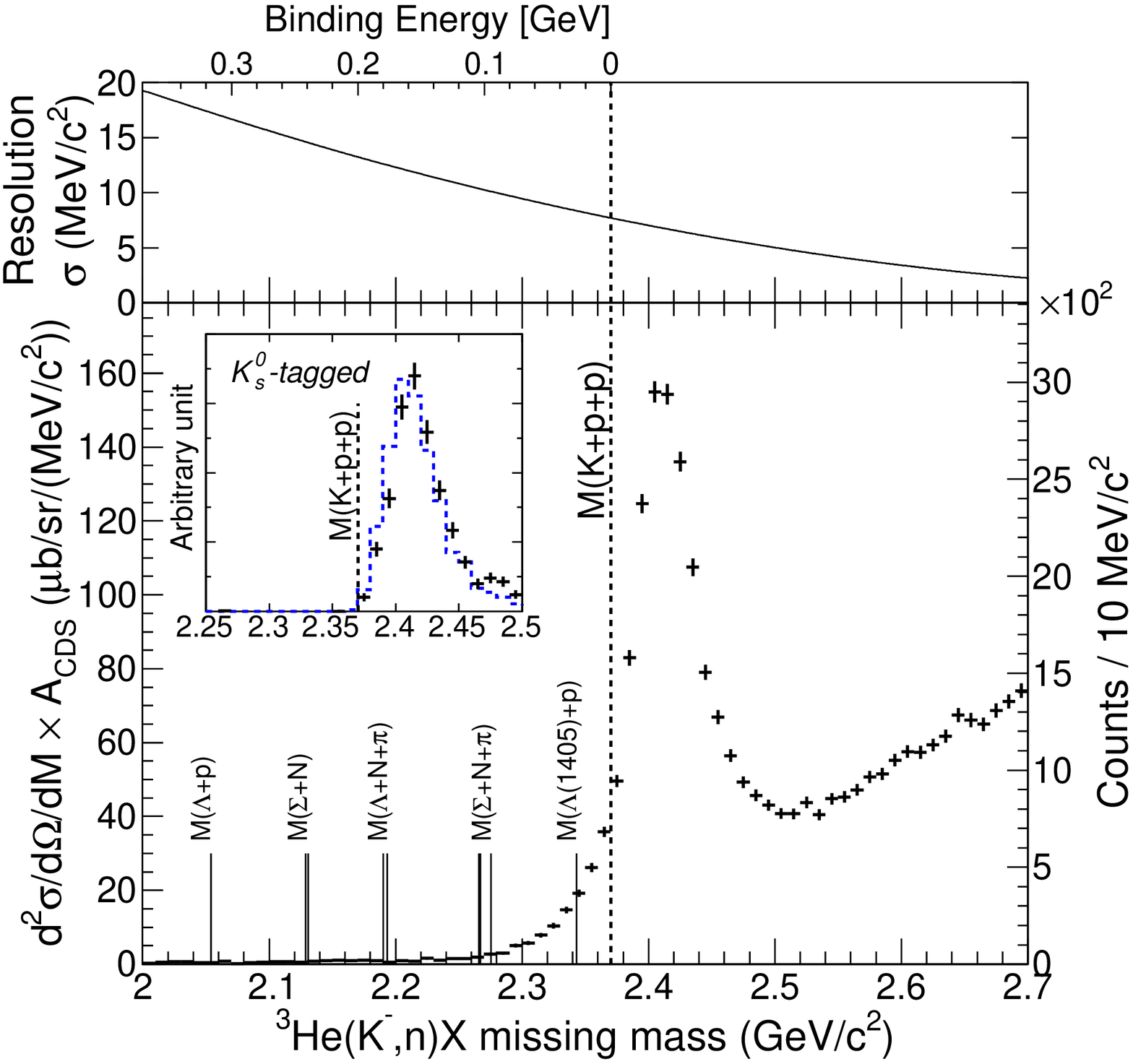}
\caption{$^3$He($K^-,n)X$ semi-inclusive missing-mass distribution below and the experimental resolution above. The $K^-pp$ binding threshold is indicated by a dotted line and other related mass thresholds are indicated by solid lines. The error bars are statistical uncertainties only. For the left-axis normalization, there is 17\% uncertainty. The inset shows the $K^0_s$-tagged spectrum compared with the simulation (blue dotted line). See Sec. 3.3 for the details of the simulation. \label{fig-mm}}
\end{center}
\end{figure}
Figure~\ref{fig-mm} shows the $^3$He($K^-,n)X$ missing-mass distribution obtained based on the $semi$-inclusive condition along with the experimental resolution. Just above the $K^-pp$ binding threshold indicated by a dotted line, the quasi-elastic peak is clearly seen. The inset of Fig.~\ref{fig-mm} shows the charge-exchange reaction by reconstructing the $K^0_s\to \pi^+\pi^-$ detected by the CDS. This $K^0_s$-tagged spectrum demonstrates that the peak broadening due to Fermi motion gives the tail structure only in the unbound region. 
The $K^0_s$-tagged spectrum is well understood with a Monte-Calro simulation taking into account the detector resolution and the Fermi motion in $^3$He \cite{Jans:1982aw}. The details of the simulation are described in Sec. 3.3.

The spectral cross section was normalized as follows:
\begin{eqnarray}
\frac{d^2\sigma}{d\Omega dM} A_{CDS} &=&\frac{1}{L}\frac{N_n}{\Omega_{NC}\Delta M\epsilon}, \nonumber
\end{eqnarray}
where $N_n$ is the number of observed neutrons in the missing-mass interval $\Delta M$, and $\Omega_{NC}$ is the acceptance of the NC. The effective integrated luminosity, $L$, obtained was 547 $\mu b^{-1}$ by considering the analysis efficiency of the kaon beam and the fiducial volume selection. The overall efficiency, $\epsilon$ = 0.16, was determined by taking into account the following factors 1) the neutron detection efficiency, 2) reaction losses between the target and the NC, 3) the vertex reconstruction efficiency, 4) the ratio of neutron over-veto, namely, real neutrons rejected by an accidental hit on the veto counter array, 5) the DAQ live rate, and 6) the trigger efficiency. $A_{CDS}$ is the CDS tagging acceptance --- namely the probability of having at least one charged particle within the CDS acceptance when a neutron is detected by the NC. There is a rather large uncertainty about $A_{CDS}$ since it depends on the angular distribution and the cross section of each reaction, which are not well known for the $K^{-3}$He reaction. Therefore, we present the spectrum without correcting for $A_{CDS}$. The systematic error of the normalization was evaluated as $\pm$ 17\%, which was dominated by the uncertainty in the neutron detection efficiency.

The quasi-elastic and charge-exchange reaction cross sections at $\theta^{lab}_n=0^\circ$ are $\sim$ 6 and $\sim$ 11 mb/sr, respectively. In the cross-section evaluation, events with a spectator deuteron or two nucleons were selected from the $^3$He($K^-,nK^0_s)$ or $^3$He($K^-,nK^-)$ reaction, and the angular distributions given in Ref. \cite{Damerell:1977wj} and \cite{Anonymous:G7Qq9mMJ} were used. The precision of the absolute missing-mass scale was also confirmed to be 3 MeV/$c^2$ from the peak position of the missing deuteron in the $^3$He($K^-,nK^0_s)X$ reaction.

\subsection{Background evaluation}
For further discussion of the spectrum obtained, four kinds of background sources are considered as follows.
\paragraph{Accidental background {\rm (${\rm BG_{accidental}}$)}}
Purely accidental background, random in time, can be evaluated using the $1/\beta$ spectrum as shown in Fig.~\ref{fig-beta}. The contribution of this background is about half of the observed yield in the energy region below the $\Lambda N$ mass threshold in the neutron missing-mass spectrum. 

\paragraph{Contamination from the target cell {\rm (${\rm BG_{cell}}$)}}
Background coming from $K^-$ reactions on the target cell, which survived even after the fiducial volume selection, were evaluated by using the empty-target data. They were coming from the finite spatial resolution of the CDS and displaced decay vertices of hyperons and $\bar K$s'.

\paragraph{Contribution from neutral particles other than neutrons {\rm (${\rm BG_{neutral}}$)}}
There is a tail component of the $\gamma$-ray peak, which comes mainly from $\pi^0$ produced via hyperon- and $\bar K$-decays. In addition, long-lived $K^0_L$ also make an NC signal through the $K^0N$ reaction and decay at the NC, appearing in the missing-mass region below the $K^-pp$ threshold. These contributions were evaluated by the Monte Carlo simulation.  

\paragraph{Fast neutrons from $\Sigma^\pm$ decays {\rm(${\rm BG_{\Sigma\mathchar`-decay}}$)}}
Background-neutron events coming from hyperon decay are not significant in the in-flight reaction. In fact, the Monte-Calro study shows fast-neutron events from $\Lambda$ decay are not triggered by the CDS. However, forward-going $\Sigma^\pm$ produced via the $K^-``N"\to \Sigma^\pm\pi$ reaction can contribute in the $K^-pp$ bound region. 
Most of these events have a CDS track of a pion coming from the $\Sigma^\pm$ decay, so that we can reconstruct the $\Sigma^{\pm}$. 
The Monte-Carlo study shows that the $\Sigma^\pm$ reconstruction efficiency for a triggered $K^-``N" \to \Sigma^{\pm}\pi$ event with a forward neutron is $\sim$ 90\% in the missing-mass region around 2.3--2.4 GeV/$c^2$, and decreases down to 70\% at around the $\Sigma+N+\pi$ mass threshold ($\sim$ 2.27 GeV/$c^2$). We have not applied the efficiency correction to ${\rm BG_{\Sigma\mathchar`-decay}}$ in the present analysis because of uncertainties from other $\Sigma^\pm$-emitting reactions, although the global structure in the $K^-pp$ bound region does not change by the correction. 

\begin{figure}
\begin{center}
\includegraphics[width=0.7\columnwidth]{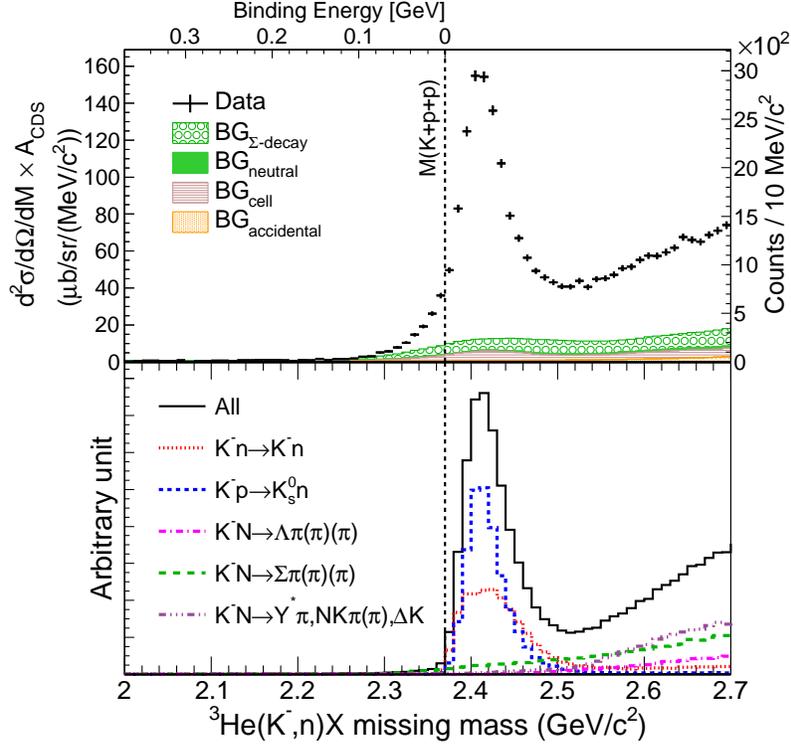}
\caption{Comparison of the $^3$He$(K^-,n)X$ semi-inclusive missing-mass spectrum between experimental data (top) and the simulation (bottom). In the experimental spectrum, the background yields are stacked on top of each other, where BG$_{\Sigma\mathchar`-decay}$ indicates $\Sigma^\pm$-reconstructed events only. In the simulation spectrum, contributions from each reaction are sorted by reaction products. \label{fig-mmsub}}
\end{center}
\end{figure}

\subsection{Comparison with the simulation of known elementary $K^-N$ reaction processes}
To understand the global structure of the spectrum obtained, known elementary $K^-N$ reaction processes were simulated based on the GEANT4 toolkit \cite{Agostinelli:2003fg}. The cross section and angular distribution of each reaction process was taken from the database \cite{Flaminio:1983uj} and the measured Fermi motion \cite{Jans:1982aw} was taken into account. 
To simulate the $^3$He target, $K^-``p"$ and $K^-``n"$ reactions were mixed with the relative cross section of 2 : 1, where half of the spectator $pn$ pairs in the $K^-``p"$ reactions are assumed to be bound as deuterons. Figure \ref{fig-mmsub} shows a comparison between the experimental data and the simulation result. The global tendency in the $K^-pp$ unbound region is well described by the $K^-N$ reactions. On the other hand, in the bound region, we observed definitive excess of the yield above the background, which cannot be explained by the present $K^-N$ simulations.

This sub-threshold structure could be attributed to the attractive $\bar K N$ interaction with an imaginary part (absorption). Another possibility is that the excess could be partially formed by the non-mesonic two-nucleon absorption processes, $K^-NN\to \Lambda(1405)n$ (or $\Sigma(1385)n$), whose energies locate just below the $K^-pp$ binding threshold. However, no other two-nucleon absorption reaction process, $K^-NN\to Y^{(*)}n$, was observed in our neutron spectrum. Therefore, we need further experimental and theoretical studies to explain why only these reaction channels are enhanced in the $K^{-3}$He reaction. 

\begin{figure}
\begin{center}
\includegraphics[width=0.7\columnwidth]{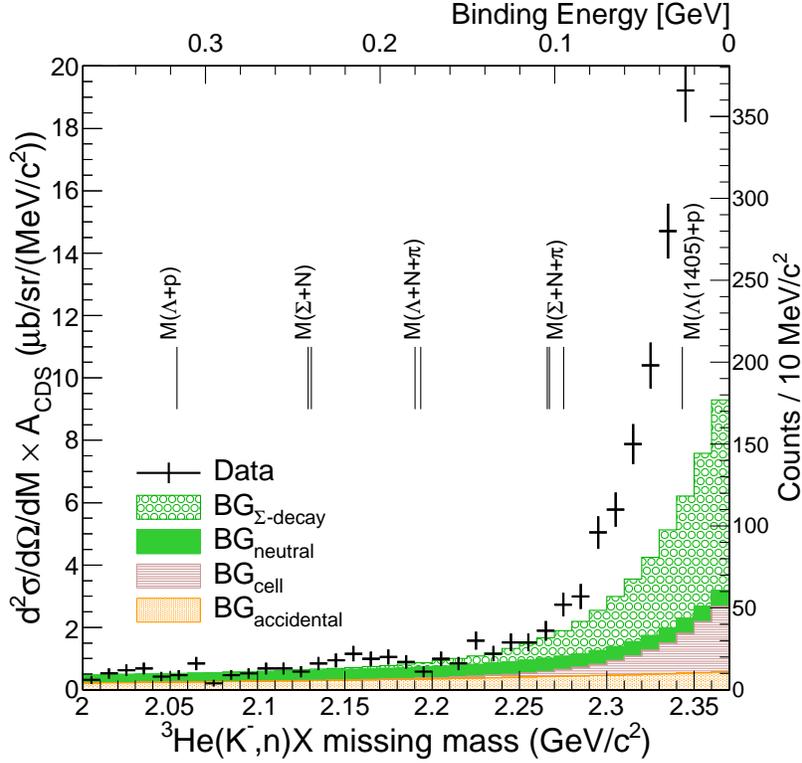}
\caption{Close-up view of $^3$He$(K^-,n)X$ semi-inclusive missing-mass spectrum (Fig.~\ref{fig-mmsub}) focusing on the $K^-pp$ bound region.  
\label{fig-mm2}}
\end{center}
\end{figure}

\subsection{Upper limits on the production cross section of a deeply bound state}
Figure~\ref{fig-mm2} shows the close-up view of the $K^-pp$ bound region. In the missing-mass region below 2.29 GeV$/c^2$, the observed events are in good agreement with the background discussed in Sec. 3.2. In the following, we focus on this deep-binding region, where the spectrum is fairly flat, thus easy to examine. Upper limits of the formation cross section for a $K^-pp$ state were determined in the mass region from $\Lambda p$ mass threshold (2.06 GeV/$c^2$) to 2.29 GeV/$c^2$, by assuming the decay branch to be $K^-pp\to \Lambda p$ in S-wave (uniform angular distribution), as suggested by FINUDA and DISTO.

The intrinsic shape of the $K^-pp$ bound state is assumed to be a Breit-Wigner function, $f(x)$. In the semi-inclusive spectrum, this intrinsic spectral shape is deformed by the CDS tagging acceptance, $A_{CDS}$, and smeared by the missing-mass resolution, $\sigma_{MM}$. The folded Breit-Wigner function, $F(x)$, is evaluated as follows:
\begin{eqnarray*}
F(x; M_X, \Gamma)&=& \int \left(f(x')\cdot A_{CDS}(x')\right)\cdot g(x-x'; \sigma_{MM}(x'))dx' ,\\
f(x; M_X, \Gamma) &=& C\cdot\left(\frac{1}{2\pi}\frac{\Gamma}{(x-M_X)^2+\Gamma^2/4}\right), \\
g(x;\sigma) &=& \frac{1}{\sqrt{2\pi}\sigma}\exp\left(-\frac{x^2}{2\sigma^2}\right), \\
\frac{d\sigma}{d\Omega}(\theta_{lab}=0) &=& \int_{M(\Lambda+p)}^{M(K^-+p+p)}f(x) dx,
\end{eqnarray*}
where $C$ is a normalization factor, and $d\sigma/d\Omega\ (\theta_{lab}=0)$, $M_X$ and $\Gamma$ are the formation cross section, mass and the natural width of the $K^-pp$ state, respectively. To evaluate $A_{CDS}$, we assumed the branching ratio of $K^-pp\to\Lambda p$ (S-wave) to be 100\%. This assumption does not affect the result substantially, because of the large CDS acceptance. In fact, $A_{CDS}$ is $\sim$ 0.7 for $K^-pp\to\Lambda p$ decay, and $\sim$ 0.4 for $K^-pp\to(\pi\Sigma)^0 p$ decay at just above the $\Sigma+p+\pi$ mass threshold ($\sim$ 2.27 GeV/$c^2$). To evaluate an upper limit on the formation cross section, a likelihood function was calculated using the function $F(x)$ with the backgrounds shown in Fig.~\ref{fig-mm2}. The systematic error, coming from the normalization factor of the cross section (17\%), was taken into account by smearing the likelihood function. Note that, in this evaluation, we used the efficiency-uncorrected $BG_{\Sigma\mathchar`-decay}$, namely $\Sigma^\pm$-reconstructed events only, hence the upper limits obtained are conservative ones.

\begin{figure}
\begin{center}
\includegraphics[width=0.7\columnwidth]{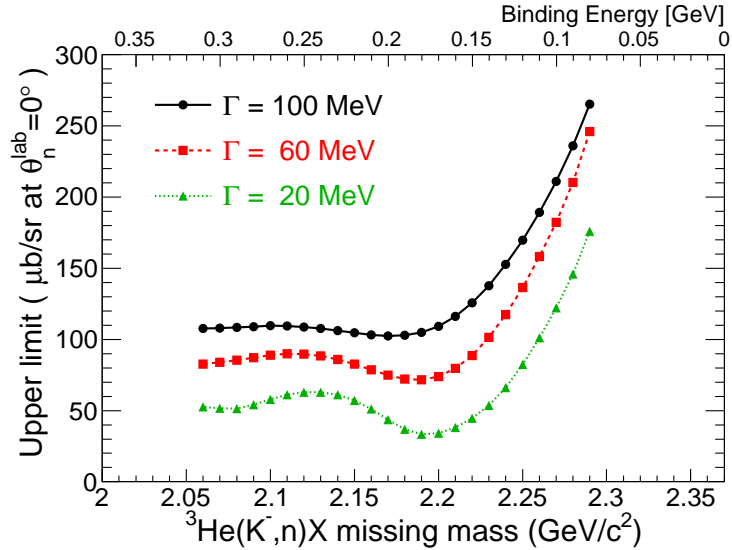}
\caption{Upper limits on the $K^-pp$ formation cross section obtained at $\theta^{lab}=0^\circ$ and 95\% confidence level for the natural widths ($\Gamma$) of 20, 60, and 100 MeV. \label{fig-mm3}}
\end{center}
\end{figure}
In Fig.~\ref{fig-mm3}, the upper limits are plotted as a function of the $^3$He($K^-,n)X$ missing mass for the natural widths of 20, 60, and 100 MeV.
The upper limits obtained are one-order-of-magnitude smaller than the theoretical calculation by Koike and Harada for the deeply-bound $K^-pp$ case. 
The ratios of the upper limits on the $K^-pp$ formation to the cross sections of the quasi-elastic channels are (0.5--5)\% (quasi-elastic $K^-``n" \to K^-n$) and (0.3--3)\% (charge exchange $K^-``p" \to K^0n$), depending on the energy and natural width. These ratios are rather small compared to the sticking probability of usual hypernucleus formation.

In the loosely-bound region, we observed a large yield below the $K^-pp$ binding threshold. The cross section of this excess above the background is $\sim$ 1 mb/sr, assuming loosely-bound $K^-pp$ formation.  
This is about the same yield as was given by Koike and Harada, but much larger than the value given by Yamagata-Sekihara {\it et al.} 
In spite of the observed large yield, the structure near the threshold suggested in the theoretical spectral functions cannot be identified from only this semi-inclusive measurement. For further investigation of the origin of the sub-threshold structure, an exclusive analysis with the forthcoming dataset is indispensable.

\section{Conclusion}
A search for the $K^-pp$ bound state was performed via the $^3$He($K^-, n)$ reaction at $\theta_{n}^{lab}=0^\circ$ at a kaon momentum of 1 GeV/$c$. In the semi-inclusive analysis, no significant peak structure was found in the $K^-pp$ deeply-bound region, where the FINUDA, DISTO and J-PARC E27 collaborations reported a bump structure in different production reactions. Mass-dependent upper limits on the production cross section were determined at 95\% confidence level in the missing-mass range from 2.06 to 2.29 GeV/$c^2$ for a $K^-pp\to\Lambda p$ isotropic decay. They were determined  to be 30--180, 70--250, and 100--270 $\mu$b/sr, for natural widths of 20, 60, and 100 MeV, respectively. These values correspond to (0.5--5)\% and (0.3--3)\% of the quasi-elastic $K^-$ and charge-exchange reaction cross sections, respectively.

\ack
We gratefully acknowledge all the staff members at J-PARC. This work was supported by RIKEN, KEK, RCNP, a Grant-in-Aid for Scientific Research on Priority Areas [No. 17070005 and No. 20028011], a Grant-in-Aid for Specially Promoted Research [No. 20002003], a Grant-in-Aid for Young Scientists (Start-up) [No. 20840047], a Grant-in-Aid for Scientific Research on Innovative Areas [No. 21105003], a Grant-in-Aid for JSPS Fellows [No. 12J10213], and the Austrian Science Fund (FWF) [21457-N16].


%

\end{document}